\documentclass{epl}
\usepackage{graphicx}
\usepackage{dcolumn}
\usepackage{bm}

\title{Dynamical mean-field equations for strongly interacting fermionic
atoms in arbitrary potential traps} \shorttitle{Dynamical mean-field
equations for trapped Fermions}
\author{W. YI \and L.-M. DUAN}
\institute{FOCUS center and MCTP, Department of Physics, University
of Michigan, Ann Arbor, Michigan 48109 USA} 
{Degenerate Fermi gases} 
\begin{document}
\maketitle

\begin{abstract}
We derive a set of general dynamical mean-field equations for
strongly interacting fermionic atoms in arbitrary slowly-varying
potential traps. Our derivation generalizes the ansatz of the
crossover wavefunction to the inhomogeneous case. The equations
reduce to a time-dependent Gross-Pitaevskii equation on the BEC side
of the resonance. We discuss an iteration method to solve these
mean-field equations, and present the solution for a harmonic trap
as an illustrating example, which self-consistently verifies the
approximations made in our derivation.
\end{abstract}

The recent experimental advance in ultracold atoms has allowed
controlled studies of strongly interacting Fermi gases in various
types of potential traps, with the interaction strength tunable by
an external magnetic field through the Feshbach resonance
\cite{1}. For homogeneous strongly interacting Fermi gases at zero
temperature, the physics is captured by the variational crossover
wavefunction \cite{2,3,4}, which interpolates the BCS and the BEC
theories. For strongly interacting atoms in a potential trap,
there are currently two main methods to deal with the resultant
inhomogeneity: one is the local density approximation (LDA)
\cite{4,5}, which neglects the kinetic terms associated with the
spatial variation of the order parameters; the other is based on
the numerical simulation of the Bogoliubov-De-Gennes (BDG)
equations \cite{6,7}. Both of these methods have found wide
applications recently \cite{4,5,6,7}, but each of them also has
its own limitation: the LDA becomes inadequate in cases where
variations of the order parameters have significant impacts; the
BDG equations take into account exactly the spatial variation of
the order-parameter, but its numerical solution is typically
time-consuming, which limits its applications only to very special
types of potentials.

In this work, we develop a different method to describe both the
static properties and the dynamics of strongly interacting fermionic
atoms in arbitrary but slowly varying potential traps. Our starting
point is a variational state which is a natural generalization of
the crossover wavefunction to the inhomogeneous and dynamical cases.
The key simplification in our derivation comes from the assumption
that the spatial variation of the order-parameter is small within
the size of the Cooper pairs. This assumption is similar to the one
in the derivation of the Ginzberg-Landau equation for the weakly
interacting fermions \cite{8}, but we avoid the use of perturbation
expansions so that the order parameter here in general does not need
to be small \cite{8,9}. With such an assumption, we derive a set of
dynamical mean-field equations for the bare molecular condensate and
the Cooper-pair wavefunctions. This set of equations can be solved
iteratively, and its zeroth-order approximation, which neglects the
order-parameter variation, gives the LDA result. On the BEC side of
the Feshbach resonance (with the chemical potential $\mu \leq 0$),
these mean-field equations can be reduced to a generalized dynamical
Gross-Pitaevskii (GP) equation \cite{8,9,castin}, with the effective
nonlinear interaction for the bare molecules derived from a
fermionic gap equation. When one goes deeper into the BEC region,
the nonlinear interaction resumes the conventional GP form, and one
can derive an effective scattering length for the bare molecules. We
solve the dynamical mean-field equations for a harmonic trap as a
simple illustrating example to self-consistently verify the
approximations made in our derivation. Another recent work also
addresses the dynamics of a trapped Fermi gas across a Feshbach
resonance \cite{added}. The set of equations derived therein are
semiclassical hydrodynamic equations, which, after linearization,
can be applied to calculate the dynamical properties of the system.
In contrast, our work follows the time-dependent variational
approach. By reducing the dynamical mean field equations to a
time-dependent non-linear GP equation on the BEC side of the
resonance, we provide a complementary perspective to the problem.

Our starting point is the two-channel field Hamiltonian \cite{3,4,11}
\begin{eqnarray}
H &=&\sum_{\sigma }\int \Psi _{\sigma }^{\dag }[-\nabla ^{2}/\left(
2m\right) +V(\mathbf{r},t)]\Psi _{\sigma }d^{3}\mathbf{r}
+\int \Psi _{b}^{\dag }[-\nabla ^{2}/\left( 4m\right) +\gamma +2V(\mathbf{r}%
,t)]\Psi _{b}d^{3}\mathbf{r}\nonumber\\
&+&\alpha \int [\Psi _{b}^{\dag }\Psi _{\downarrow }\Psi _{\uparrow }d^{3}%
\mathbf{r}+h.c.]+U\int \Psi _{\uparrow }^{\dag }\Psi _{\downarrow
}^{\dag }\Psi _{\downarrow }\Psi _{\uparrow }d^{3}\mathbf{r},
\end{eqnarray}
which describes the interaction between the fermionic atom fields
$\Psi _{\sigma }^{\dag }$ ($\sigma =\uparrow ,\downarrow $ labels
atomic internal states) in the open channel and the bosonic bare
molecule field $\Psi _{b}^{\dag }$ in the closed channel. In this
Hamiltonian, $m$ is the atom mass, and $V(\mathbf{r},t)$ is the trap
potential which could vary both in space and in time. Note that we
have assumed that the trap frequencies for a composite boson and for
a single atom are the same, so that the potential that a boson feels
is twice as a single atom does. The bare atom-molecule coupling rate
$\alpha $, the bare background scattering rate $U$, and the bare
energy detuning of the closed channel molecular level relative to
the threshold of the two-atom continuum $\gamma $ are connected with
the physical ones $\alpha _{p},U_{p},\gamma _{p}$ through the
standard renormalization relations \cite{4}. The values of the
physical parameters $\alpha _{p},U_{p},\gamma _{p}$ are determined
respectively from the resonance width, the background scattering
length, and the magnetic field detuning relative to the Feshbach
resonance point (see, e.g., the explicit expressions in Ref.
\cite{d1}). Note that following the standard two-channel model,
direct collisions between the bosonic bare molecules are neglected,
as their contribution is negligible near a broad Feshbach resonance
\cite{3,4,11}.

At almost zero temperature and with a slowly varying potential $V(\mathbf{r}%
,t)$, the state of the Hamiltonian (1) can be assumed to evolve according to
the following variational form:
\begin{equation}
|\Phi \left( t\right) \rangle =\mathcal{N} \exp \left[ \int f(\mathbf{r},%
\mathbf{r}^{\prime },t)\Psi _{\uparrow }^{\dag }(\mathbf{r})\Psi
_{\downarrow }^{\dag }(\mathbf{r}^{\prime })d^{3}\mathbf{r}d^{3}\mathbf{%
r^{\prime }}\right]
 \exp \left[ \int \phi _{b}(\mathbf{r},t)\Psi _{b}^{\dag }(\mathbf{r}%
)d^{3}\mathbf{r}\right] |vac\rangle ,
\end{equation}
where $\mathcal{N}$ is the normalization factor, $\phi _{b}(\mathbf{r},t)$
is the condensate wavefunction for the bare molecules, and $f(\mathbf{r},%
\mathbf{r}^{\prime },t)$ is the Cooper-pair wavefunction. This variational
state is a natural generalization of the crossover wavefunction to the
inhomogeneous and dynamical cases \cite{yd}. Without the fermionic field,
this variational state would have the same form as the one in the derivation
of the dynamical GP equation for the weakly interacting bosons \cite{castin}.

To derive the evolution equations for the wavefunctions $\phi _{b}(\mathbf{r}%
,t)$ and $f(\mathbf{r},\mathbf{r}^{\prime },t)$, we follow the standard
variational procedure to minimize the action $S=\int dt[\langle \dot{\Phi}%
|\Phi \rangle -\langle \Phi |\dot{\Phi}\rangle ]/(2i)-\langle \Phi |H|\Phi
\rangle $, where $\left| \Phi \right\rangle $ and $H$ are specified in Eqs.
(1) and (2). Under the ansatz state (2), the Wick's theorem implies the
decomposition $U\langle \Psi _{\uparrow }^{\dag }\Psi _{\downarrow }^{\dag
}\Psi _{\downarrow }\Psi _{\uparrow }\rangle \approx U\langle \Psi
_{\uparrow }^{\dag }\Psi _{\downarrow }^{\dag }\rangle \langle \Psi
_{\downarrow }\Psi _{\uparrow }\rangle $ (the additional Hartree-Fock terms,
which only slightly modify the effective $V(\mathbf{r},t)$, are not
important when there is pairing instability \cite{8} and are thus neglected
here). It turns out that to get the expression of the action $S$, the
critical part is the calculation of the pair function $F^{\ast }(\mathbf{%
r_{1}},\mathbf{r_{2}},t)\equiv \langle \Psi _{\uparrow }^{\dag }(\mathbf{%
r_{1}})\Psi _{\downarrow }^{\dag }(\mathbf{r_{2}})\rangle $. Under the
ansatz state (2), the pair function satisfies the following integral
equation (we drop the time variables in $F^{\ast }(\mathbf{r_{1}},\mathbf{%
r_{2}},t)$ and $f^{\ast }(\mathbf{r},\mathbf{r}^{\prime },t)$ when there is
no confusion)
\begin{equation}
F^{\ast }(\mathbf{r_{1}},\mathbf{r_{2}}) =f^{\ast }(\mathbf{r_{1}},\mathbf{%
r_{2}})
-\int f^{\ast }(\mathbf{r_{1}},\mathbf{r_{3}})f^{\ast }(\mathbf{r_{4}},%
\mathbf{r_{2}})F(\mathbf{r_{4}},\mathbf{r_{3}})d^{3}\mathbf{r}_{3}d^{3}%
\mathbf{r}_{4}.
\end{equation}
To solve this integral equation, we write both $F^{\ast }(\mathbf{r_{1}},%
\mathbf{r_{2}})$ and $f^{\ast }(\mathbf{r_{1}},\mathbf{r_{2}})$ in terms of
the new coordinates $\mathbf{r}=(\mathbf{r_{1}}+\mathbf{r_{2}})/2$ and $%
\mathbf{r}_{-}=\mathbf{r_{1}}-\mathbf{r_{2}}$. Then, we take the Fourier
transformation of Eq. (3) and its conjugate with respect to the relative
coordinate $\mathbf{r}_{-}$. The Fourier transforms of $F(\mathbf{r},\mathbf{%
r_{-}})$ and $f(\mathbf{r},\mathbf{r_{-}})$ are denoted by $F_{\mathbf{k}}(%
\mathbf{r})$ and $f_{\mathbf{k}}(\mathbf{r})$, respectively. In this Fourier
transformation, we assume $\left| \partial _{\mathbf{r}}f\right| \ll \left|
\partial _{\mathbf{r}_{-}}f\right| $ and $\left| \partial _{\mathbf{r}%
}F\right| \ll \left| \partial _{\mathbf{r}_{-}}F\right| $. Physically, it
corresponds to the assumption that the order parameter is slowly varying
within the size of the Cooper pairs. Under this assumption, we derive from
Eq. (3) and its conjugate the following simple relation between the Fourier
components
\begin{equation}
F_{\mathbf{k}}(\mathbf{r})=f_{\mathbf{k}}(\mathbf{r})/\left[ 1+|f_{\mathbf{k}%
}(\mathbf{r})|^{2}\right] .
\end{equation}
This relation is critical for the explicit calculation of the action $S$.

We can now express the action $S$ in terms of the variational wavefunctions $%
f_{\mathbf{k}}(\mathbf{r},t)$ and for $\phi _{b}(\mathbf{r},t)$. From the
functional derivatives $\delta S/\delta f_{\mathbf{k}}^{\ast }(\mathbf{r}%
,t)=0$ and $\delta S/\delta \phi _{b}^{\ast }(\mathbf{r},t)=0$, we get the
following evolution equations for $f_{\mathbf{k}}(\mathbf{r},t)$ and for $%
\phi _{b}(\mathbf{r},t)$:
\begin{eqnarray}
i\partial _{t}f_{\mathbf{k}} &=&[2\epsilon _{\mathbf{k}}+H_{0}(\mathbf{r}%
,t)]f_{\mathbf{k}}+\Delta (\mathbf{r},t)-\Delta ^{\ast }(\mathbf{r},t)f_{%
\mathbf{k}}^{2},\hspace{0.5cm} \\
i\partial _{t}\phi _{b} &=&[\gamma +H_{0}(\mathbf{r},t)]\phi _{b}+\left(
\alpha /U\right) \Delta _{f}(\mathbf{r},t),\hspace{0.5cm}
\end{eqnarray}
where $H_{0}(\mathbf{r},t)\equiv -\nabla ^{2}/\left( 4m\right)
+2V(\mathbf{r} ,t)$, $\epsilon_{\mathbf{k}}=\hbar^2\mathbf{k}^2/2m$
($m$ is the atomic mass), $\Delta _{f}(\mathbf{r},t)\equiv \left(
U/8\pi ^{3}\right) \int d^{3}
\mathbf{k}f_{\mathbf{k}}(\mathbf{r},t)/\left[
1+|f_{\mathbf{k}}(\mathbf{r} ,t)|^{2}\right] $, and $\Delta
(\mathbf{r},t)\equiv \alpha \phi _{b}(\mathbf{ r},t)+\Delta
_{f}(\mathbf{r},t)$. The two equations (5) and (6) represent a
central result of this work: they completely determine the evolution
of the wavefunctions $f_{\mathbf{k}}$ and for $\phi _{b}$, just as
the GP equation determines the condensate evolution for the weakly
interacting bosons. In the stationary case with a time-independent
trap, one just needs to replace $ i\partial _{t}f_{\mathbf{k}}$ and
$i\partial _{t}\phi _{b}$ respectively with $2\mu f_{\mathbf{k}}$
and $2\mu \phi _{b}$, where $\mu $ is the atom chemical potential.

The evolution equations (5) and (6) are a set of coupled nonlinear
differential equations. They can be solved through direct numerical
simulations (for instance, through the split-step method), but as the
potential $V(\mathbf{r},t)$ is typically slowly varying both in $\mathbf{r}$
and in $t$, the following iterative method may prove to be more efficient.
In this case, we expect both $\phi _{b}(\mathbf{r},t)e^{i2\mu t}$ and $f_{%
\mathbf{k}}(\mathbf{r},t)e^{i2\mu t}$ to be slowly varying in $\mathbf{r}$
and $t$. We can then introduce the following effective potentials $V_{eff}(%
\mathbf{r},t)\equiv \phi _{b}^{-1}(\mathbf{r},t)\left[ -i\partial
_{t}-\nabla ^{2}/\left( 4m\right) +2\mu \right] \phi _{b}(\mathbf{r},t)/2$
and $V_{eff}^{\mathbf{k}}(\mathbf{r},t)\equiv f_{\mathbf{k}}^{-1}(\mathbf{r}%
,t)\left[ -i\partial _{t}-\nabla ^{2}/\left( 4m\right) +2\mu \right] f_{%
\mathbf{k}}(\mathbf{r},t)/2$, both of which should be small. With these
introduced potentials, we can solve $f_{\mathbf{k}}(\mathbf{r},t)$ from Eq.
(5) as
\begin{equation}
f_{\mathbf{k}}=-(E_{\mathbf{k}}-(\epsilon _{\mathbf{k}}-\mu _{eff}^{\mathbf{k%
}}))/\Delta ^{\ast }
\end{equation}
where $\mu _{eff}^{\mathbf{k}}\equiv \mu -V(\mathbf{r},t)-V_{eff}^{\mathbf{k}%
}(\mathbf{r},t)$, $\mu _{eff}\equiv \mu -V(\mathbf{r},t)-V_{eff}(\mathbf{r}%
,t)$, $\Delta =\alpha \phi _{b}[1-U(\gamma -2\mu _{eff})/\alpha ^{2}]$, and $%
E_{\mathbf{k}}=\sqrt{(\epsilon _{\mathbf{k}}-\mu _{eff}^{\mathbf{k}%
})^{2}+\Delta ^{2}}$. Substituting Eq. (7) into Eq. (6), we get the
following effective gap equation
\begin{equation}
1/U_{T}=-\left( 1/8\pi ^{3}\right) \int d^{3}\mathbf{k}(\frac{1}{2E_{\mathbf{%
k}}}-\frac{1}{2\epsilon _{\mathbf{k}}}),
\end{equation}
where $1/U_{T}=1/\left[ U_{p}-\alpha _{p}^{2}/\left( \gamma
_{p}-2\mu _{eff}\right) \right] =1/\left[ U-\alpha ^{2}/\left(
\gamma -2\mu _{eff}\right) \right] +\int d^{3}\mathbf{k}\left[
1/\left( 16\pi ^{3}\epsilon _{\mathbf{k}}\right) \right] $ (the
latter equality comes from the renormalization relation between
$\gamma ,\alpha ,U$ and $\gamma _{p},\alpha _{p},U_{p}$ \cite{4}).

Under the zeroth-order approximation, we assume $V_{eff}(\mathbf{r},t)\simeq
V_{eff}^{\mathbf{k}}(\mathbf{r},t)\simeq 0$, which leads to $\mu _{eff}^{%
\mathbf{k}}=\mu _{eff}=\mu -V(\mathbf{r},t)$ in $E_{\mathbf{k}}$. In this
case, the gap equation (8), together with the number equation $N=\int n(%
\mathbf{r},t)d^{3}\mathbf{r}$, where $N$ denotes the total atom number and $%
n(\mathbf{r},t)=2|\phi _{b}|^{2}+\left( 2/8\pi ^{3}\right) \int d^{3}\mathbf{%
k}|f_{\mathbf{k}}|^{2}/(1+|f_{\mathbf{k}}|^{2})$ is the local atom density,
completely solves the problem, the result of which corresponds to a solution
under the local density approximation in the adiabatic limit. Thus we
recover the LDA result under the zeroth-order approximation which completely
neglects $V_{eff}(\mathbf{r},t)$ and $V_{eff}^{\mathbf{k}}(\mathbf{r},t)$.
It is then evident as how to go beyond the LDA. We can use the LDA result $%
\phi _{b}^{\left( 0\right) }(\mathbf{r},t),f_{\mathbf{k}}^{\left( 0\right) }(%
\mathbf{r},t)$ as the zeroth-order wavefunctions to calculate the
first-order effective potentials $V_{eff}^{\left( 1\right) }(\mathbf{r},t)$
and $V_{eff}^{\mathbf{k}\left( 1\right) }(\mathbf{r},t)$ through their
definition equations. Substituting these effective potentials into the gap
equation (8) and (7), we can find out the next order wavefunctions $\phi
_{b}^{\left( 1\right) }(\mathbf{r},t),f_{\mathbf{k}}^{\left( 1\right) }(%
\mathbf{r},t)$. This iterative process should converge if the effective
potentials are small (i.e., the order parameters are slowly varying in $%
\mathbf{r}$ and $t$).

In the following, we consider a different simplification of the basic
equations (5) and (6) on the BEC side of the resonance with the chemical
potential $\mu \leq 0$ (note that it is not required to be in the deep BEC
region). On this side, we expect the wavefunctions $\phi _{b}(\mathbf{r},t)$
and $f_{\mathbf{k}}(\mathbf{r},t)$ to have similar dependence on $\mathbf{r}$
and $t$, so we assume $V_{eff}(\mathbf{r},t)\simeq V_{eff}^{\mathbf{k}}(%
\mathbf{r},t)$. This approximation will be self-consistently tested and we
will see that it is well satisfied when $\mu \leq 0$. Under this
approximation, $\mu _{eff}^{\mathbf{k}}=\mu _{eff}$, while $\mu _{eff}$ can
be solved from the gap equation (8) as a function of $\left| \phi
_{b}\right| ^{2}$. Substituting this solution $\mu _{eff}$ into the
definition equation of $V_{eff}$, we get
\begin{equation}
i\partial _{t}\phi _{b}=\left[ -\nabla ^{2}/\left( 4m\right) +2V(\mathbf{r}%
,t)+2\mu _{eff}\right] \phi _{b}.
\end{equation}
This equation has the same form as the dynamical GP equation for the weakly
interacting bosons except that the collision term is now replaced by a
general nonlinear potential $\mu _{eff}$ \cite{castin}, a function of $%
\left| \phi _{b}\right| ^{2}$ with its shape determined by the gap
equation (8). We have numerically solved Eq.(8) for the function
$\mu _{eff}(\left| \phi _{b}\right| ^{2})$ at several different
detunings for $^{6}$Li, and the shapes of these functions are shown
in Fig. 1. We can see that $\mu _{eff}$ becomes almost linear in
$|\phi _{b}|^{2}$ when one goes further into the BEC region. In that
limit, Eq. (9) reduces to an exact GP equation, and we can define an
effective scattering length $a_{eff}$ for the bare molecular
condensate with $d\mu _{eff}/d(|\phi _{b}|^{2})=2\pi a_{eff}/\left(
2m\right) $. This effective scattering length $a_{eff}$ is shown in
Fig. 1(d) as a function of the field detuning for $^{6}$Li. We
should note, however, that the effective scattering length for the
bare molecules is in general very different from the one for the
dressed molecules \cite{4,9,pe}. The dressed molecules are
dominantly composed of Cooper pairs of atoms in the open channel
(for instance, when the chemical potential $\mu \approx 0$,
the population fraction of the bare molecules is only about $0.1\%$ for $%
^{6}$Li). As the bare molecules have a very low density near the
resonance, they in general have a large effective scattering length
to compensate for that, as is shown in Fig. 1(d). The effective
scattering lengths for the bare and the dressed molecules coincide
with each other only in the deep BEC region with the population
dominantly in the closed channel. In this limit, we have checked
that the dependence of the effective scattering length on the field
detuning is in agreement with a different calculation in Ref. \cite
{12} under the two-channel model (we can only apply the two-channel
model in this limit because of a large closed channel population
\cite{4,13}). Experimentally, the scattering length between the
dressed molecules can be measured from the collective excitations of
the trapped Fermi gas \cite{14} or from the in-trap radius of the
condensate cloud \cite{15}; while it is difficult to measure the
scattering length between the bare molecules.

\begin{figure}[tbp]
\onefigure{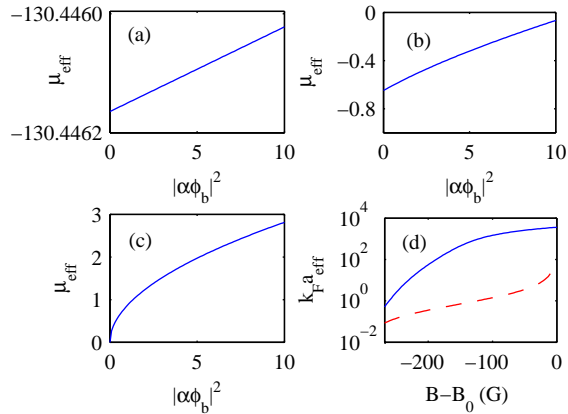} \caption[Fig.1 ]{(a)(b)(c) The effective
potential $\protect\mu _{eff}$
as a function of $|\protect%
\alpha \protect\phi _{b}|^{2}$, with the field detuning $B-B_{0}$
given by $-268G$ (a), $-107G$ (b), and $0G$ (c), respectively.
Both $\protect\mu _{eff}$ and $|\protect%
\alpha \protect\phi _{b}|$ are in the unit of $E_F=k_{F}^{2}/2m$,
where $k_{F}=(3\pi ^{2}n_{0})^{1/3}$ is a convenient inverse length
scale corresponding to a density $n_{0}=3\times 10^{13}cm^{-3}$ as
it is typical for the MIT ${}^{6}$Li experiment \cite{1}. (d) The
effective scattering lengths as a function of the field detuning.
The solid line is for the bare molecules (see the definition in the
text) while the dashed one is for the atoms. }
\end{figure}

The simplified equation (9) determines the distribution of the
bosonic molecules. This solution, combined with Eq. (7), then
fixes the distribution of the fermionic atoms. As a simple
illustrating example, we use them to solve the fermi condensate
shape function in a harmonic trap on the BEC side
of the resonance. We take the values of the parameters corresponding to $%
^{6}$Li, and assume a total of $N\sim 10^{5}$ atoms trapped in a
time-independent potential $V(\mathbf{r})=\frac{1}{2}m\omega
\mathbf{r}^{2}$ with $\omega /2\pi \sim 100Hz$, as is typical in the
experiments \cite{1}. Figures 2(a) and 2(b) show the condensate
shape functions in two different
regions with the magnetic field detunings $B-B_{0}$ given respectively by $%
-268G$ and $-107G$. The first detuning corresponds to a point deep in the
BEC region with $(k_{F}a_{s})^{-1}\sim 11$, where $a_{s}$ is the atom
scattering length at that detuning and $k_{F}^{-1}$ is a convenient length
unit defined in the caption of Fig. 1. The second one corresponds to the
onset of the bosonic region with the atom chemical potential $\mu \sim 0$
and $(k_{F}a_{s})^{-1}\sim 0.8$. We have shown in Fig. 2 the number
distributions for the bare molecules and the fermi condensate. One can see
that these distribution functions are smooth in space, without the
artificial cutoff at the edge of the trap as in the LDA result. The closed
channel population (the total bare molecule fraction) is calculated to be
about $33\%$ and $0.1\%$ respectively for these two detunings.

\begin{figure}[tbp]
\onefigure{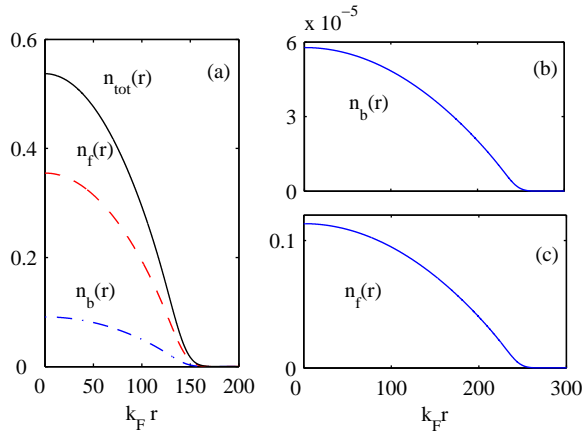} \caption[Fig.2 ]{The radial distributions
of of the bare-molecule density ($n_{b}$), the fermi-atom density
($n_{f}$), and the total density ($n_{tot}$) in a harmonic trap with
the field detuning $B-B_0$ given by $-268G$ (a), and $-107G$ (b and
c), respectively. All the densities are in the unit of $n_0$ (see
the value in the caption of Fig. 1)}
\end{figure}

An important goal of calculation of this simple example is to self
consistently check the approximations made in our derivation. First,
to derive the basic equations (5,6), we have assumed that the order
parameter should be slowly varying over the size of the Cooper
pairs. From Fig. (2), we see that the characteristic length for the
variation of the order parameter is typically of $100k_{F}^{-1}$,
while the size of the Cooper pair is well below $k_{F}^{-1}$ at
these detunings \cite{13}. Therefore, this approximation should be
well satisfied for typical experiments. Second, from the basic
equations (5,6) to the simplified equation (9), we have used the
approximation $V_{eff}(\mathbf{r},t)\simeq
V_{eff}^{\mathbf{k}}(\mathbf{r},t) $. To check the validity of this
approximation, we calculate the effective potentials
$V_{eff}(\mathbf{r})$ and $V_{eff}^{\mathbf{k}}(\mathbf{r})$
(time-independent in this case) with our solutions of $\phi
_{b}(\mathbf{r})$ and $f_{\mathbf{k}}(\mathbf{r})$ from the
stationary harmonic trap, and the
results are shown in Fig. 3. It is clear that the difference $\left| V_{eff}(%
\mathbf{r})-V_{eff}^{\mathbf{k}}(\mathbf{r})\right| $ is small compared with
the magnitude of $\left| V_{eff}(\mathbf{r})\right| $ for different values
of $\mathbf{k}$ when the atom chemical potential $\mu \leq 0$, which
justifies the approximation $V_{eff}^{\mathbf{k}}(\mathbf{r},t)\simeq
V_{eff}(\mathbf{r},t)$ in that region. One can also see that the relative
error $\left| V_{eff}(\mathbf{r})-V_{eff}^{\mathbf{k}}(\mathbf{r})\right|
/\left| V_{eff}(\mathbf{r})\right| $ goes up significantly (from roughly $%
10^{-4}$ to $10^{-1}$) when one goes from the field detuning $-268G$ to $%
-107G$. If one goes further to the resonance point, this approximation
eventually breaks down, and one needs to use the basic equations (5,6)
instead of the reduced equation (9).

\begin{figure}[tbp]
\onefigure{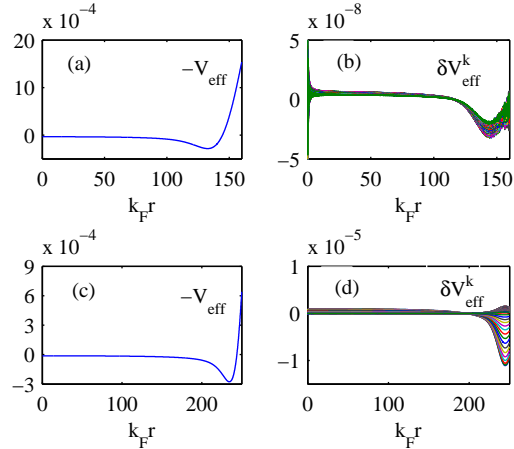} \caption[Fig.3 ]{The shapes of the
effective potential $-V_{eff}$ and the potential difference $\delta
V_{eff}^{\mathbf{k}}=V_{eff}^{\mathbf{k}}-V_{eff}$ (both in the unit
of $E_F$) at the field detuning $B-B_0=-268G$ (a,b) and $-107G$
(c,d), respectively. Figs. (a) and (d) each have twenty curves
corresponding to the wave vector $|\mathbf{k}|$ varies from $0$ to
$10k_F$.}
\end{figure}

In summary, we have derived a set of dynamical mean-field equations for
evolution of strongly interacting fermionic atoms in any slowly varying
potential traps, and discussed methods to solve these equations. We show
that on the BEC side of the Feshbach resonance, this set of equations are
reduced to a generalized dynamical GP equation. As an illustrating example,
we solve the reduced equations in the case of a harmonic trap, which
self-consistently verifies the approximations made in our derivation.

This work was supported by the NSF award (0431476), the ARDA under ARO
contracts, and the A. P. Sloan Fellowship.

\end{document}